\documentclass[pra,reprint,twocolumn,amsmath,amssymb,superscriptaddress]{revtex4-1}
\usepackage{graphicx}
\usepackage{physics}
\usepackage{siunitx}
\usepackage{mathtools}
\usepackage{float}
\usepackage{bm}
\usepackage{bbm}
\usepackage[dvipsnames, table]{xcolor}
\usepackage[colorlinks=true,citecolor=Cerulean,linkcolor=RubineRed,urlcolor=Cerulean]{hyperref}
\usepackage[T1]{fontenc}
\usepackage{letltxmacro}

\LetLtxMacro{\originaleqref}{\eqref}
\renewcommand{\eqref}[1]{Eq.~(\ref{#1})}
\newcommand{\figref}[1]{Fig.~\ref{#1}}

\newcommand{\up}{{\ket{\uparrow}}}
\newcommand{\down}{{\ket{\downarrow}}}

\newcommand{\fleft}{\mathopen{}\mathclose\bgroup\left}
\newcommand{\fright}{\aftergroup\egroup\right}

\begin{document}
\title{Spin-Dependent Momentum Conservation of Electron-Phonon Scattering in Chirality-Induced Spin Selectivity}

\author{Clemens Vittmann}
\affiliation{Institut f\"ur Theoretische Physik und IQST, Albert-Einstein-Allee 11, Universit\"at Ulm, D-89081 Ulm, Germany}
\author{James Lim}
\affiliation{Institut f\"ur Theoretische Physik und IQST, Albert-Einstein-Allee 11, Universit\"at Ulm, D-89081 Ulm, Germany}
\author{Dario Tamascelli}
\affiliation{Institut f\"ur Theoretische Physik und IQST, Albert-Einstein-Allee 11, Universit\"at Ulm, D-89081 Ulm, Germany}
\affiliation{Dipartimento di Fisica ``Aldo Pontremoli'', Universit{\`a} degli Studi di Milano, Via Celoria 16, 20133 Milano-Italy}
\author{Susana F. Huelga}
\affiliation{Institut f\"ur Theoretische Physik und IQST, Albert-Einstein-Allee 11, Universit\"at Ulm, D-89081 Ulm, Germany}
\author{Martin B. Plenio}\email{martin.plenio@uni-ulm.de}
\affiliation{Institut f\"ur Theoretische Physik und IQST, Albert-Einstein-Allee 11, Universit\"at Ulm, D-89081 Ulm, Germany}

\begin{abstract}
The elucidation of the mechanisms behind chiral-induced spin selectivity remains an outstanding scientific challenge. Here we consider the role of delocalised phonon modes in electron transport in chiral structures and demonstrate that spin selectivity  can originate from spin-dependent energy and momentum conservation in electron-phonon scattering events. While this mechanism is robust to the specifical nature of the vibrational modes, the degree of spin polarization depends on environmental factors, such as external driving fields, temperatures and phonon relaxation rates. This dependence is used to present experimentally testable predictions of our model.
\end{abstract}

\maketitle

{\it Introduction.} The capability of chiral molecules to polarize electron spins, called chirality-induced spin selectivity (CISS)~\cite{FirstCISS,EnantiospecificSpinPol_2014,naaman_spintronics_2015,kettner_2018,aiello2020chiralitybased}, has been observed in a variety of organic molecules such as DNA~\cite{Goehler894}, bacteriorhodopsine~\cite{mishra_spin-dependent_2013}, photosystem~I~\cite{carmeli_spin_2014} or oligopeptides~\cite{kettner_spin_2015}. The electron spin filtering effect has been demonstrated for both the transmission of photoelectrons through chiral media~\cite{Goehler894,mishra_spin-dependent_2013,kettner_spin_2015}, such as monolayers of DNA~\cite{Goehler894}, and the transport of bound electrons through chiral molecules between electrodes~\cite{mishra_spin-dependent_2013,carmeli_spin_2014,kettner_spin_2015}. More recently, the interplay of electron spin and enantioselectivity in chiral molecules has been the subject of increased attention~\cite{kumar_chirality-induced_2017, banerhee-ghosh_2018, metzger_electron_2020, dianat_2020, kapon_evidence_2021,CISSVanDerWaals}. These experimental findings have initiated the development of chirality-based spintronic nanodevices and technologies~\cite{bostick_2018, dor_chiral-based_2013, Michaeli_2017, mondal_spin-dependent_2016, yang_2020, Chiesa2021}. Although the spin-orbit coupling induced by chiral molecules has been considered a key parameter to explain the high spin polarization observed in experiments~\cite{Cuniberti2012,Guo_2012,GutierrezCunibertiEffective,guo_spin-dependent_2014,geyer_chirality-induced_2019,naaman_chiral_2019,Naaman_Michaeli_2019,fransson_chirality-induced_2019,dalum_theory_2019,GhazaryanAnalyticModel2020,geyer_2020,ZhangPRB2022phonon,VibEnhanced,Fransson2020,FranssonNL2021,TempDependentCISS2022}, there is not yet a consensus on the microscopic mechanism of CISS~\cite{Overview2021}. In this work, we go beyond these models to demonstrate that non-equilibrium dynamics of {\em delocalised phonon modes} of chiral molecules induce fluctuations in spin orbit couplings which, in combination with the principle of {\em spin-dependent energy and momentum conservation}, can provide a mechanism for spin selectivity.

%%%%%%%%%%%%%%% Figure
\begin{figure}[t]
	\includegraphics[width=0.45\textwidth]{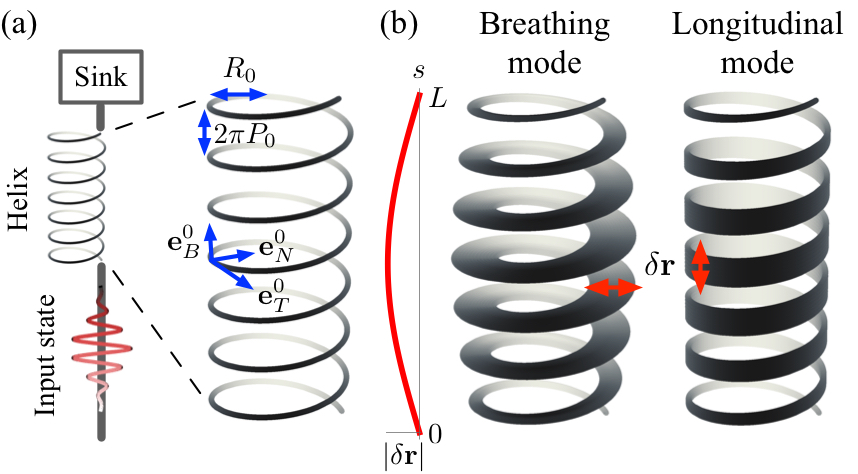}
	\caption{(a) A schematic representation of a single-stranded helix coupled to input and output electrodes. (b) Fundamental breathing and longitudinal modes where the degree $|\delta{\mathbf r}(s)|$ of structural deformation, indicated by red arrows, depends on position $s$.}
	\label{Fig1}
\end{figure}

{\it Model.} We consider a single-stranded helix coupled to electrodes, as shown in \figref{Fig1}(a), where an initial spin-unpolarized Gaussian electron wavepacket propagates from input, via the helix, to output electrode. In the absence of phonon motion, the helical electron path of a length $L$ is parameterized by $\mathbf r_0(s)=(R_0\cos(s/D),R_0\sin(s/D),P_0 s/D)$ with $0\le s \le L$ and $D=\sqrt{R_0^2+P_0^2}$, where $R_0$ and $2\pi P_0$ denote, respectively, its radius and pitch. The electron propagates in the tangential direction of the helical path determined by the unit vector ${\bf e}_{T}^{0}(s)= \partial_s {\mathbf r}_0/|\partial_s {\mathbf r}_0|$. In conventional CISS models~\cite{Cuniberti2012,Guo_2012,GutierrezCunibertiEffective,guo_spin-dependent_2014}, the electric field acting on the electron wavepacket, induced by surrounding charges, is assumed to be helically symmetric and oriented in a normal direction described by ${\bf e}_{N}^{0}(s)=-(\cos(s/D),\sin(s/D),0)$. A position $s$-dependent orthonormal basis is formed by considering an additional binormal unit vector, defined by ${\bf e}_{B}^{0}(s)={\bf e}_{T}^{0}(s)\times {\bf e}_{N}^{0}(s)$. With electron momentum oriented in the tangential direction, ${\bf p}={\bf e}_{T}^{0}(s)(-i\hbar\partial_s)$, the electron Hamiltonian has been modeled by
\begin{equation}
    H = \frac{p_s^2}{2m} + \frac{\alpha_0}{2}\{\bm\sigma\cdot\bm e^0_B(s),p_s\},
    \label{eq:Hstatic}
\end{equation}
where $p_s=-i\hbar\partial_s$, $m$ denotes electron mass and $\alpha_0$ quantifies the spin-orbit coupling strength. The last term originates from symmetrization of a spin-orbit coupling in the form $\boldsymbol{\sigma}\cdot ({\bf E}\times{\bf p})$ where $\boldsymbol{\sigma}$ are the Pauli matrices, while ${\bf E}=E(R(s)){\bf e}_{N}^{0}(s)$ represents the electric field whose amplitude $E(R(s))$ depends on the radius $R(s)$ in the polar coordinate where the helix is aligned in the $z$-direction. It has been shown that such a single-stranded helix does not show an electron-spin filtering effect when it is coupled to the electrodes via single-point contacts~\cite{Guo_2012, Overview2021}, and a finite interface between helix and electrodes is essential to observe a CISS effect in simulations~\cite{OurPaper}. In the present work, we consider single-point contacts to demonstrate that coupling of electrons to delocalized phonons can induce non-zero spin polarization even if the interfacial effects are absent.

A deformation $\delta\mathbf r(s)$ of the unperturbed helical electron path, $\bm r_0(s)$, to $\bm r(s) = \bm r_0(s) + \delta\mathbf 
r(s)$ due to phonons induces perturbations in the electric field acting on the electron wavepacket, $\delta{\bf E}$, and in the orientation of electron momentum, $\delta{\bf p}$, as its propagation direction deviates from the tangential direction of the unperturbed helical path (${\bm e}^0_T(s)\neq\partial_{s}{\bf r}/|\partial_{s}{\bf r}|$). This results in phonon-induced fluctuations of the spin-orbit coupling $\boldsymbol{\sigma}\cdot (({\bf E}+\delta{\bf E})\times({\bf p}+\delta{\bf p}))$, where $\boldsymbol{\sigma}\cdot ({\bf E}\times{\bf p})$ is a static component considered in conventional CISS models, while the strength of the remaining dynamic components depends on the magnitude of the phonon motion.

For simplicity, we consider delocalized breathing and longitudinal motion of a helix, inducing position-dependent fluctuations of the radius $R(s)$ and pitch $2\pi P(s)$, respectively. Here the degree of fluctuations may depend on the position $s$ along the helical path, as shown in \figref{Fig1}(b), as determined by phonon wavelengths. 

In the presence of the delocalized breathing motion, the helical path is deformed in the normal direction, $\delta \bm r(s)=\bm e^0_N(s) \delta r(s)$ with $\delta r(s)=\sum_{k}A_{k}(t)\sin(\beta_{k} s)$. Here $\hbar\beta_{k}$ and $A_{k}(t)$ represent, respectively, the momenta and time-dependent classical amplitudes of delocalized phonon modes with wavelengths $2\pi\beta_{k}^{-1}$. For small deformations, the electric field acting on the electron wavepacket can be expanded to first order in $\delta r(s)$ which yields ${\bf E}\approx(E(R_0)+\partial_{R}E|_{R=R_0}\delta r(s)){\bf e}_{N}^{0}(s)$ and thus $\delta{\bf E}=\partial_{R}E|_{R=R_0}\delta r(s){\bf e}_{N}^{0}(s)$. Similarly, the orientation ${\bm e}_{T}(s)=\partial_s\bm r/|\partial_s\bm r|$ of electron momentum operator in the perturbed helix may deviate slightly from the tangential direction ${\bm e}_{T}^{0}(s)$ of the unperturbed helical path and includes normal and binormal components ${\bm e}_{N,B}^{0}(s)$. With the modified orientation of the electron momentum operator, ${\bf p}+\delta{\bf p}={\bm e}_{T}(s)(-i\hbar\partial_s)$, the total spin orbit coupling is described by
\begin{align}
    \frac{\alpha_0}{2}\{ \bm\sigma\cdot ({\bm e}_{T}(s) \cross (1+\lambda \delta r(s))\bm e^{0}_{N}(s)), p_s\},
    \label{eq:breathing_v0}
\end{align}
where $p_s=-i\hbar\partial_s$ and $\lambda=\partial_{R}E|_{R=R_0}/E(R_0)$. As shown in the SI, the dynamic components of the spin-orbit coupling in Eq.~(\ref{eq:breathing_v0}) are expressed as
\begin{align}
    \sum_{j\in\{B,T\}}\frac{\alpha_{j}}{2} \{\boldsymbol{\sigma}\cdot{\bm e^{0}_{j}}(s)\sum_k A_k(t)\sin(\beta_k s),p_s\},
    \label{eq:breathing}
\end{align}
where $\alpha_{B}=\alpha_0 (\lambda-R_0 D^{-2})$ and $\alpha_{T}=-\alpha_0 P_0 D^{-2}$. 

Similarly, the dynamic spin-orbit couplings can be derived for the longitudinal motion where the helical electron path is deformed in the $z$-direction, $\delta\bm r(s) = {\bm e_z}\delta r(s)$. As shown in the SI, the dynamic spin-orbit couplings are expressed as
\begin{align}
    \sum_{j\in\{B,T\}}\frac{\alpha'_{j}}{2} \{\boldsymbol{\sigma}\cdot{\bm e^{0}_{j}}(s)\sum_k A_k(t)\beta_k\cos(\beta_k s),p_s\},
    \label{eq:longitudinal}
\end{align}
where $\alpha'_{B}=\alpha_0P_0 D^{-1}$ and $\alpha'_{T}=-\alpha_0R_0 D^{-1}$.

In this work, we consider quantized phonon modes whose energies are described by the Hamiltonian of quantum harmonic oscillators $H_v=\sum_{k}\hbar\omega_k (a_{k}^{\dagger}a_{k}+1/2)$ where $a_k^{\dagger}$ and $a_k$ denote, respectively, the creation and annihilation operators of a phonon mode with frequency $\omega_k$. The dynamic spin-orbit couplings induced by the quantized phonon modes are obtained by replacing classical phonon amplitudes $A_k(t)$ with $(a_k+a_{k}^{\dagger})$ and a suitable prefactor, which are reduced to linear combinations of
\begin{align}
    H_1&=\frac{\alpha}{2}\sum_{k}(a_{k}+a_{k}^{\dagger})\{\sigma(s)\sin(\beta_k s),p_s\}\quad\text{or}\label{eq:H1}\\
    H_2&=\frac{\alpha}{2}\sum_{k}(a_{k}+a_{k}^{\dagger})\{\sigma(s)\beta_k\cos(\beta_k s),p_s\},\label{eq:H2}
\end{align}
where $\sigma(s) =\boldsymbol{\sigma}\cdot{\bm e^{0}_{B/T}}(s)$ and $\alpha$ quantifies the overall dynamic spin-orbit coupling strength. Here the spin operators are represented in the $\sigma_z$ eigenbasis by
\begin{align}
    \boldsymbol{\sigma}\cdot{\bm e^{0}_{B}}(s)&=\frac{1}{D}
    \begin{pmatrix}
        R_0 & iP_0 e^{-is/D} \\
        -iP_0 e^{is/D} & -R_0
    \end{pmatrix},\label{eq:sigma_B}\\
    \boldsymbol{\sigma}\cdot{\bm e^{0}_{T}}(s)&=\frac{1}{D}
    \begin{pmatrix}
        P_0 & -iR_0 e^{-is/D} \\
        iR_0 e^{is/D} & -P_0
    \end{pmatrix}.\label{eq:sigma_T}
\end{align}
We stress that the spin-flip coupling from $\up$ to $\down$ and that from $\down$ to $\up$ carry opposite position $s$-dependent phase factors $e^{is/D}$ and $e^{-is/D}$, respectively. In the remainder of this work, we will demonstrate that these opposite phases form a key ingredient of our mechanism for phonon-induced spin selectivity. Note also that this mechanism does not depend on details of the dynamic spin-orbit coupling, which may be $\boldsymbol{\sigma}\cdot{\bm e^{0}_{B}}(s)$ or $\boldsymbol{\sigma}\cdot{\bm e^{0}_{T}}(s)$, shown separately in Eqs.~(\ref{eq:H1}--\ref{eq:H2}), or their linear combinations, shown in Eqs.~(\ref{eq:breathing}--\ref{eq:longitudinal}), describing breathing and longitudinal modes. In this work, we will consider $\boldsymbol{\sigma}\cdot{\bm e^{0}_{B}}(s)$ terms in Eq.~(\ref{eq:H1}) for simplicity, as the other forms lead to a similar dependence of spin polarization on phonon parameters.

\begin{figure}[t]
	\includegraphics[width=0.48\textwidth]{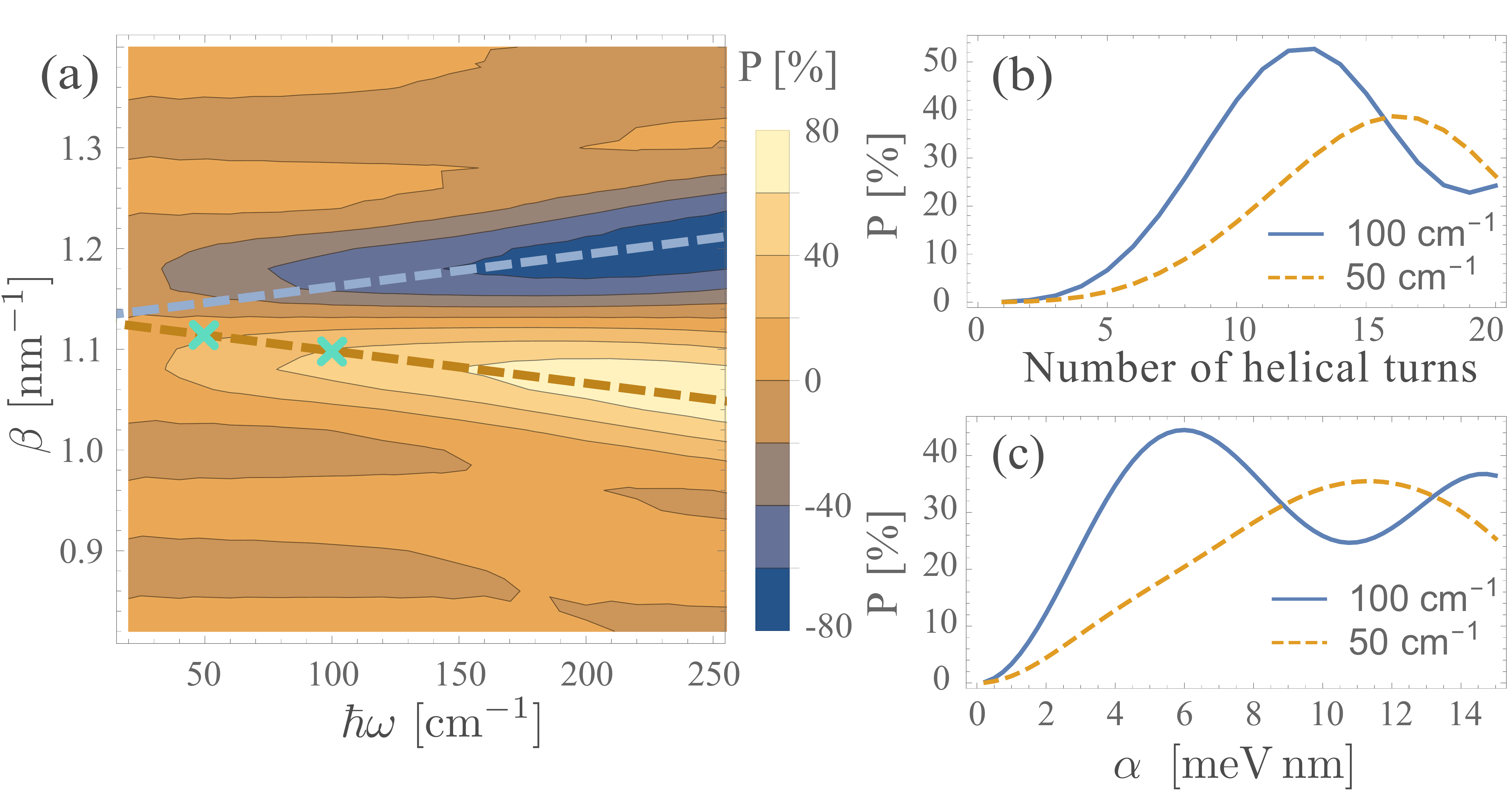}
	\caption{(a) Spin polarization as a function of phonon frequency $\omega$ and momentum $\hbar\beta$. The phonon mode is initially in its vacuum state and the initial electron wavepacket has a kinetic energy of $\SI{1}\electronvolt$. A broad initial Gaussian wavepacket is considered so that its average momentum is an order of magnitude larger than the standard deviation of its momentum distribution (see the SI for more details). Here we consider $\alpha=\SI{5}{\milli\electronvolt\nano\meter}$ and ten helical turns $L/(2\pi D)=10$. The helix is modeled by radius $R_0=\SI{0.7}{\nm}$ and pitch $2\pi P_0=\SI{3.4}{\nm}$, leading to $D^{-1}=\SI{1.130}{\per\nm}$. Two dashed lines mark solutions to energy-momentum conservation conditions (see the main text). For the two sets of phonon parameters marked in (a), where $\hbar\omega\in\{50,100\}\,{\rm cm}^{-1}$, the spin polarization is plotted as a function of (b) the number $L/(2\pi D)$ of helical turns and (c) the dynamic spin-orbit coupling strength $\alpha$ with other parameters as in (a).}
	\label{Fig2}
\end{figure}

{\it Spin-dependent momentum conservation of electron-phonon scattering.} To identify the influence of the electron-phonon interaction on electron spin polarization, we consider a single phonon mode with frequency $\omega$ and momentum $\hbar\beta$, modeled by $H=H_0 + H_{d}$ where $H_0=p_s^2/(2m) + \hbar\omega(a^\dagger a+1/2)$ describes the energy of electron and phonon mode, while $H_{d}=(\alpha/2)(a+a^\dagger)\{\boldsymbol{\sigma}\cdot{\bm e^{0}_{B}}(s)\sin(\beta s),p_s\}$ is the dynamic spin-orbit coupling from Eq.~(\ref{eq:H1}). The eigenstates of $H_0$ are described by $\ket{k,S,n}=(2\pi\hbar)^{-1/2}\int\dd se^{iks}\ket{s,S,n}$ where $\ket{k}$ denote electron's momentum eigenstates, $\ket{S}$ with $S\in\{\uparrow,\downarrow\}$ the spin eigenstates of the $\sigma_z$ operator, and $\ket{n}$ phonon number states with $n$ being non-negative integers. The dynamic spin-orbit coupling $H_{d}$ induces the transitions between $\ket{k,S,n}$, including both spin-flipping and spin-conserving processes. For instance, we may consider a spin-up initial state $\ket{k,\uparrow,0}$ where the phonon mode is in its vacuum state $\ket{0}$. The transition from the initial state to a spin-flipped final state $\ket{k',\downarrow,1}$ may satisfy the energy conservation condition $(\hbar^2/2m)(k^2-k'^2)=\hbar\omega$, where the electron momentum $\hbar k'$ after an electron-phonon scattering event is determined by the input electron momentum $\hbar k$ and the energy quanta $\hbar\omega$ of the phonon mode. For given $\Delta k = k-k'$, the corresponding transition amplitude mediated by the dynamic spin-orbit coupling $H_{d}$ is expressed as
\begin{equation}
    \bra{k',\downarrow,1}H_{d}\ket{k,\uparrow,0}\propto \int_{0}^{L}ds e^{is/D}\sin(\beta s)e^{i\Delta k s},
\end{equation}
whose magnitude increases as $|\Delta k + (D^{-1} \pm\beta)|$ becomes smaller. This is contrary to the transition amplitude from an initially spin-down state $\ket{k,\downarrow,0}$ to a final spin-up state $\ket{k',\uparrow,1}$ whose magnitude is enhanced as $|\Delta k + (-D^{-1}\pm\beta)|$ becomes smaller. Since we are considering delocalized phonon modes, typically having low phonon frequencies $\hbar \omega \lesssim 100\,{\rm cm}^{-1}$ in case of helical molecules such as DNA and alpha helix~\cite{PhononsInDNA}, the change in electron momentum is small and therefore the spin-flip dynamics is governed by the following spin-dependent momentum conservation condition
\begin{align}
    \label{eq:EnergyMomentumCons}
    \up\rightarrow\down:&\quad k-\sqrt{k^2-(2m/\hbar)\omega}=-D^{-1}+\beta,\\ \label{eq:EnergyMomentumCons2}
    \down\rightarrow\up:&\quad k-\sqrt{k^2-(2m/\hbar)\omega}=D^{-1}-\beta.
\end{align}
The momentum conservation condition depends on $D^{-1}+\beta$ only when $|\Delta k|$ is sufficiently large, requiring a high phonon frequency $\omega$. These results indicate that negative and positive spin polarizations can be observed when $\beta>D^{-1}$ and $\beta<D^{-1}$, respectively, in the zero-temperature limit where the phonon mode is initially in its vacuum state (see Fig.~\ref{Fig2}(a)). Importantly, the energy and momentum conservation conditions are identical for all the dynamic spin-orbit coupling forms shown in Eqs.~(\ref{eq:breathing}--\ref{eq:H2}), as the spin-dependence originates from the phase factors $e^{\pm i s/D}$ of the spin-flipping terms in Eqs.~(\ref{eq:sigma_B}--\ref{eq:sigma_T}).

%%%%%%%%%%%%%%% Figure

To demonstrate that Eqs.~(\ref{eq:EnergyMomentumCons}--\ref{eq:EnergyMomentumCons2}) governs spin dynamics of our model, the initial electron state is taken to be a random mixture of spin-up and down states whose spatial profile is described by a Gaussian wavepacket entering from the input electrode (see \figref{Fig1}(a) and the SI for more details). For simplicity, the Hamiltonian is modeled by $H=H_0+\Pi(s)H_{d}$ where $\Pi(s)$ is a rectangular function defined by $\Pi(s)=1$ if $0\le s\le L$ and $\Pi(s)=0$ otherwise, describing the electron-phonon interaction $H_{d}$ within a helix. To avoid an unrealistic sudden change in Hamiltonian, we consider a smoothstep function at the interfaces between electrodes and helix on a sub-nm length scale in simulations (see the SI), rounding off the edges of the rectangular function. As a consequence, the electron wavepacket enters the helix and then exits through the output electrode with a negligible back reflection to the input electrode. In simulations, the electron wave packet dynamics is computed by using a finite difference method until the electron wavepacket is fully transferred to the output electrode after an electron-phonon scattering event. A sink, which we call Markovian closure~\cite{TamascelliPRL2018,MC2022}, is introduced in the output electrode (see Fig.~\ref{Fig1}(a)) to prevent re-entering of the wavepacket from the output electrode to the helix, induced by a finite length of the electrode considered in simulations. The sink does not affect spin polarization of the final electron wavepacket, but makes simulations more stable and efficient (see the SI for more details).

For the zero temperature limit, \figref{Fig2}(a) shows the final spin polarization as a function of phonon frequency $\omega$ and momentum $\hbar\beta$, where negative and positive spin polarizations appear along dashed lines where the energy and momentum conservation conditions in Eqs.~(\ref{eq:EnergyMomentumCons}--\ref{eq:EnergyMomentumCons2}) are satisfied. \figref{Fig2}(b) and (c) show that the final spin polarization oscillates as a function of the overall dynamic spin-orbit coupling strength $\alpha$ and the length $L$ of a helix, and the spin polarization can reach several tens of percents for a wide parameter range and moderate spin-orbit coupling. These results demonstrate that the phonon-assisted spin-dependent momentum conservation can induce a high electron spin polarization.

{\it External driving, thermal noise and relaxation of phonon modes.} So far we have considered an undamped phonon mode that is initially in its vacuum state and decoupled from any environments. Here we consider various scenarios where the phonon mode is pumped by external infrared fields and/or relaxes to a thermal equilibrium state due to the interaction with environments. An important goal of this discussion is the identification of effects that may be amenable to experimental test.

\begin{figure}[t]
	\includegraphics[width=0.48\textwidth]{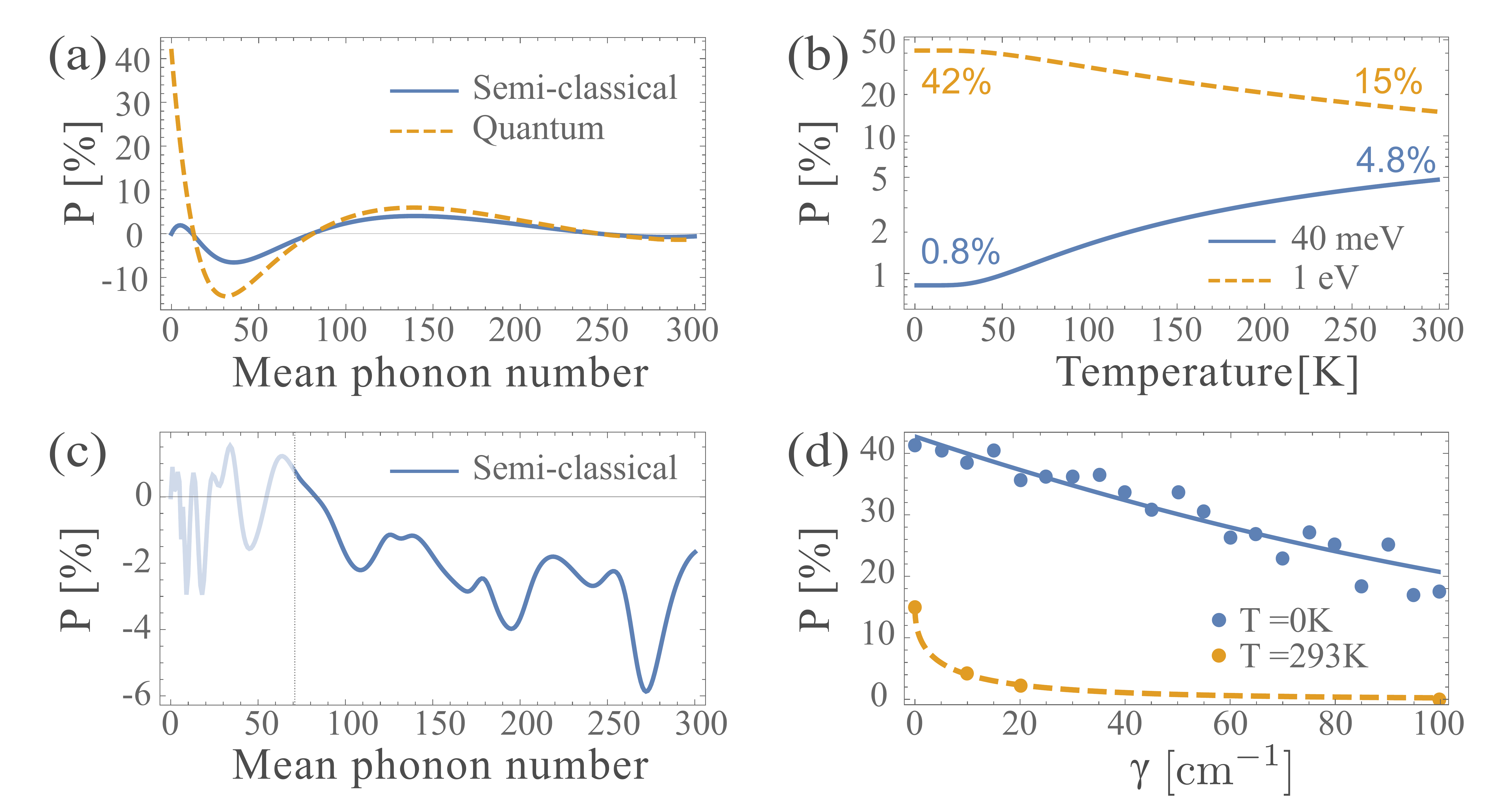}
	\caption{(a) Spin polarization as a function of mean phonon number $|\xi|^{2}$ of initial coherent state $\ket{\xi}$, computed by a full quantum and by a semi-classical model, for an initial electronic kinetic energy of $1\,{\rm eV}$. (b) Temperature dependence of spin polarization where a phonon mode is initially in a thermal state with initial electronic kinetic energy of $1\,{\rm eV}$ or $40\,{\rm meV}$. In (a,b), a single delocalized phonon mode is considered with $\omega=100\,{\rm cm}^{-1}$, $\beta\approx 1.098\,{\rm nm}^{-1}$ and other parameters as in \figref{Fig2}(a) unless stated otherwise. (c) Spin polarization as a function of $|\xi|^{2}$, computed by the semi-classical model of part (a) where 100 classical phonon modes with amplitudes $\xi e^{-i\omega_k t}$ are considered with constant $\xi$ and phonon frequencies $\omega_k$ uniformly distributed between 1 and $100\,{\rm cm}^{-1}$. Here the momentum $\hbar\beta_k$ of every phonon mode satisfies spin-dependent energy and momentum conservation conditions (see dashed line in \figref{Fig2}(a) where $\omega\in\{50,100\}\,{\rm cm}^{-1}$ points are marked) and uniform dynamic spin-orbit coupling strengths $\alpha_k=0.4\,{\rm meV\,nm}$ are considered. (d) Spin polarization as a function of phonon relaxation rate $\gamma$ at zero and room temperatures where model parameters are as in (a).}
	\label{Fig3}
\end{figure}

In \figref{Fig3}(a), the phonon mode is assumed to be initialized in a coherent state $\ket{\xi}=\sum_{n=0}^{\infty}e^{-|\xi|^{2}/2}(\xi^{n}/\sqrt{n!})\ket{n}$ with a mean phonon number $|\xi|^2$, e.g. by an external infrared laser pulse. The resulting spin polarization depends strongly on the mean phonon number, suggesting that CISS effects supported by delocalized phonon modes could be tested experimentally by varying the intensity of externally applied laser pulses. We have also computed the spin polarization using a semi-classical model where the phonon mode operator $a$ is replaced by a time-dependent classical phonon amplitude $\xi e^{-i\omega t}$. It is found that the predictions of the semi-classical model are qualitatively similar to those of the full quantum model based on a quantized phonon mode, but only approach the full model quantitatively when the mean phonon number is sufficiently large, $|\xi|^{2} \gtrsim 70$. The difference between model predictions for small $|\xi|^{2}$, including the case of the vacuum state with $\xi=0$, demonstrates that phonon-induced CISS effects cannot be fully described by the semi-classical approach. In \figref{Fig3}(b), we consider a thermal state of the delocalized phonon mode at the initial time, which is a statistical mixture of the phonon number states $\ket{n}$ with the Boltzmann distributions being proportional to $e^{-\hbar\omega n/k_B T}$. Interestingly it is found that depending on the initial kinetic energy of the electron wavepacket, the spin polarization can be decreased or increased as a function of temperature $T$. This implies that various temperature dependences could be observed when the injection electron energy is controlled, for instance, by tuning the properties of electrodes and the voltage between them.

%%%%%%%%%%%%%%% Figure

So far we have considered a single delocalized phonon mode. To take into account the situation that the electron wavepacket is coupled to multiple delocalized phonon modes, we consider two different approaches. In \figref{Fig3}(c), we use the semi-classical model where 100 phonon modes with frequencies $\omega_k$ uniformly distributed between 1 and $100\,{\rm cm}^{-1}$ are coupled to the electron wavepacket. For $\xi$ independent of the mode frequency and concentrating on the regime for which the semiclassical model becomes quantitative, $|\xi|^2 \gtrsim 70$, the spin polarization is found to grow in magnitude with the mean phonon number $|\xi|^2$, suggesting, yet again, the possibility that vibrational CISS effects can be controlled by external infrared fields driving multiple phonon modes. Finally, in \figref{Fig3}(d), we consider a delocalized phonon mode coupled to a thermal environment. Initially in a thermal state any perturbation of the phonon mode due to an electron-phonon scattering can then relax back towards the thermal state thanks to the interaction with its environment. The phonon relaxation is modeled by a Lindblad master equation
\begin{align}
    \partial_t \rho(t) &= -(i/\hbar)[H,\rho(t)] +\gamma n(a^{\dagger}\rho(t)a-\{a a^{\dagger}\!,\rho(t)\}/2)\nonumber\\
    &\quad + \gamma (1+n)(a\rho(t)a^{\dagger}
    -\{a^{\dagger}a,\rho(t)\}/2),
\end{align}
with $\rho(t)$ and $n=(\exp(\hbar\omega/k_B T)-1)^{-1}$ representing, respectively, a density matrix of a hybrid electron-phonon system and the average number of thermal excitations at temperature $T$. Here the effective phonon mode under Lindbladian dynamics emulates a phonon bath with a continuous frequency spectrum centered at $\omega$ with a width proportional to the phonon relaxation rate $\gamma$~\cite{TamascelliPRL2018,SomozaPRL2019}. By using a quantum jump algorithm~\cite{PlenioRMP1988} (see the SI for more details), spin polarization is observed for moderate values of $\gamma\lesssim \omega$ at both zero and room temperatures. These results demonstrate that phonon-induced CISS mechanisms could be robust against environmental effects.

{\it Conclusions.} We have demonstrated that delocalized phonon modes can result in spin-dependent momentum conservation of electron-phonon scattering events, which, in turn, provides a mechanism for high spin polarization in helical structures. This phonon-induced CISS effect depends on various environmental factors, such as external driving fields, temperature and phonon relaxation rate. On the one hand, this implies that vibrational CISS effects can be tested in experiments by controlling the state of phonon modes, e.g. as we propose by application of infrared laser pulses or, as recently explored in Ref.~\cite{TempDependentCISS2022}, by temperature changes. On the other hand, our results show that microscopic modeling of chiral molecules is important to identify the origin of CISS as the degree of spin polarization strongly depends on phonon parameters. It is notable that non-equilibrium phonon dynamics was considered explicitly in our simulations, contrary to independent theoretical studies where classical phonon modes were considered~\cite{ZhangPRB2022phonon} or quantized phonon modes were assumed to instantaneously relax to their thermal equilibrium states during electron-phonon interaction~\cite{VibEnhanced,Fransson2020,FranssonNL2021,TempDependentCISS2022}. These approximate treatments of phonon modes cannot describe the spin-dependent energy and momentum conservation conditions, governing spin dynamics in our model, and their reliability at finite temperatures remains an open question. A more detailed analysis of our model and its extension, including both delocalized and localized phonon modes as well as consequent spin-dependent current-voltage characteristics, deserves a separate investigation and will be presented in a forthcoming manuscript.

\begin{acknowledgments}
This work was supported by the ERC Synergy grant HyperQ (Grant No.\ 856432). The authors acknowledge support by the state of Baden-Württemberg through bwHPC and the German Research Foundation (DFG) through Grant No.\ INST 40/575-1 FUGG (JUSTUS~2 cluster).
\end{acknowledgments}

\newpage

\begin{widetext}

\

\begin{center}
{\large\bf Supporting Information}
\end{center}

\section{Derivation of Dynamic Spin-Orbit Couplings}
\subsection{Breathing Motion}

Here we derive the dynamic spin-orbit coupling induced by breathing motion. As explained in the main manuscript, a helical electron path in the presence of the breathing motion is described by
\begin{align}
    \bm{r}(s) = \bm{r}_0(s) + \delta\bm{r}(s),
\end{align}
where $\bm{r}_0(s)$ represents an unperturbed helical path with radius $R_0$ and pitch $2\pi P_0$
\begin{align}
    \bm{r}_0(s) =
    \begin{pmatrix}
        D\cos(\chi)\cos(\frac{s}{D}) \\ D\cos(\chi)\sin(\frac{s}{D})\\\sin(\chi) s
    \end{pmatrix},
\end{align}
while $\delta\bm{r}(s)$ denotes the phonon-induced deformation in the normal direction
\begin{align}
    \delta\bm{r}(s) = \bm{e}^0_N(s) \sum_k A_k(t)\sin(\beta_k s).
\end{align}
Here $\tan(\chi) = \tau_0/\kappa_0$ with $\tau_0=P_0/D^2$ and $\kappa_0=R_0/D^2$ representing, respectively, the torsion and curvature of the unperturbed helical path with $D=\sqrt{R_0^2+P_0^2}$. $\{\bm e^0_T(s),\bm e^0_N(s), \bm e^0_B(s)\}$ denote the Frenet-Serret vectors, forming an orthonormal basis depending on position $s$, defined by
\begin{align}
    {\bm e}^{0}_{T}(s) =\begin{pmatrix} 
    -\cos(\chi)\sin(\frac{s}{D}) \\ \cos(\chi)\cos(\frac{s}{D}) \\ \sin(\chi)
    \end{pmatrix},~
    {\bm e}^{0}_{N}(s) =\begin{pmatrix} 
    -\cos(\frac{s}{D}) \\ -\sin(\frac{s}{D}) \\ 0
    \end{pmatrix},~
    {\bm e}^{0}_{B}(s) =\begin{pmatrix} 
    \sin(\chi)\sin(\frac{s}{D}) \\ -\sin(\chi)\cos(\frac{s}{D}) \\ \cos(\chi)
    \end{pmatrix}.
\end{align}
For the perturbed helical path, the orientation of electron momentum is described by ${\bm e}_T(s)=\partial_s {\bm r}/|\partial_s {\bm r}|$ where
\begin{align}
    \partial_s {\bm r} &= \bm e_T^0(s) + (\sin(\chi) \bm e_B^0(s) - \cos(\chi) \bm e_T^0(s)) \frac{\delta r(s)}{D} + \bm e_N^0(s) \partial_s \delta r(s),
\end{align}
with $\delta r(s)=\sum_k A_k(t)\sin(\beta_k s)$ and $\partial_s \delta r(s)=\sum_k A_k(t)\beta_k \cos(\beta_k s)$. When the electric field is approximately described by ${\bm E}\approx (E(R_0)+\partial_R E|_{R=R_0}\delta r(s)){\bm e}^{0}_{N}(s)$, as discussed in the main manuscript, the total spin-orbit coupling is expressed as
\begin{align}
    H_{\rm SOC}=\frac{\alpha_0}{2}\{{\bm \sigma}\cdot ({\bm e}_{T}(s)\times (1+\lambda \delta r(s)){\bm e}^{0}_{N}(s)),p_s\},
\end{align}
where $\lambda = \partial_R E|_{R=R_0}/E(R_0)$ and $p_s = -i\hbar\partial_s$. When the deformation $\delta_r(s)$ is sufficiently small, $|\partial_s {\bm r}| \approx 1$ as $\delta r(s)/D\ll 1$ and $\partial_s \delta r(s)\ll 1$. In this case, with ${\bm e}^{0}_T(s)\times {\bm e}^{0}_N(s)={\bm e}^{0}_B(s)$ and ${\bm e}^{0}_B(s)\times {\bm e}^{0}_N(s)=-{\bm e}^{0}_T(s)$, the total spin-orbit coupling up to first order in $\delta r(s)$ is given by
\begin{align}
    H_{\rm SOC}=\frac{\alpha_0}{2}\{{\bm \sigma}\cdot{\bm e}_{B}(s),p_s\}+\left(\frac{\alpha_0(\lambda-\cos(\chi)/D)}{2}\{{\bm \sigma}\cdot{\bm e}_{B}(s),p_s\}-\frac{\alpha_0(\sin(\chi)/D)}{2}\{{\bm \sigma}\cdot{\bm e}_{T}(s),p_s\}\right)\delta r(s),
\end{align}
where $\cos(\chi)/D=R_0/D^{2}$ and $\sin(\chi)/D=P_0/D^2$. The first term is a static spin-orbit coupling, which has been considered in conventional CISS models, while the other terms describe dynamic spin-orbit couplings whose strengths depend on $\delta r(s)$.

\subsection{Longitudinal Motion}

Here we derive the dynamic spin-orbit coupling induced by longitudinal modes. In the presense of the longitudinal motion, the perturbation ${\bm r}(s)$ of a helical electron path is described by
\begin{align}
    \delta{\bm r}(s)={\bm e}_{z}\sum_{k}A_k(t)\sin(\beta_k s),
\end{align}
with ${\bm e}_{z}$ denoting a unit vector in the $z$-direction. In this case, the orientation ${\bm e}_{T}(s)=\partial_s {\bm r}/|\partial_s {\bm r}|$ of electron momentum is determined by
\begin{align}
    \partial {\bm r} = {\bm e}^{0}_{T}(s)+{\bm e}_{z}\partial_s \delta r(s) = {\bm e}^{0}_{T}(s)+(\sin(\chi) {\bm e}_{T}^{0}(s) + \cos(\chi) {\bm e}_{B}^{0}(s))\partial_s \delta r(s),
\end{align}
with $\partial_s \delta r(s)=\sum_{k}A_k(t) \beta_k \cos(\beta_k s)$. Since the electric field is assumed to be invariant under the translation in the $z$-direction in our model (see the main manuscript), in the limit of a small perturbation $\delta r(s)$, leading to $|\partial_s {\bm r}|\approx 1$, the total spin-orbit coupling is given by
\begin{align}
    H_{\rm SOC}&=\frac{\alpha_0}{2}\{{\bm \sigma}\cdot ({\bm e}_{T}(s)\times {\bm e}^{0}_{N}(s)),p_s\}\\
    &=\frac{\alpha_0}{2}\{{\bm \sigma}\cdot{\bm e}_{B}(s),p_s\}+\left(\frac{\alpha_0 \sin(\chi)}{2}\{{\bm \sigma}\cdot{\bm e}_{B}(s),p_s\}-\frac{\alpha_0\cos(\chi)}{2}\{{\bm \sigma}\cdot{\bm e}_{T}(s),p_s\}\right)\partial_s \delta r(s),
\end{align}
where $\sin(\chi)=P_0/D$ and $\cos(\chi)=R_0/D$. Here the first term is a static spin-orbit coupling, while the remaining terms describe dynamic spin-orbit couplings whose strengths depend on $\partial_s \delta r(s)$.

\section{Transition Amplitude of Spin-Flipping Processes}

In the main manuscript, we consider the Hamiltonian $H=H_0+H_d$ with $H_0=p_{s}^{2}/(2m)+\hbar\omega(a^\dagger a+1/2)$ and $H_d=(\alpha/2)(a+a^\dagger)\{{\bm \sigma}\cdot {\bm e}_{B}^{0}(s)\sin(\beta s),p_s\}$. The transition amplitude between spin-up state $\ket{k,\uparrow,0}$ and spin-down state $\ket{k',\downarrow,1}$ mediated by $H_d$ is given by
\begin{align}
    \bra{k',\downarrow,1} H_{d} \ket{k,\uparrow,0}&=\frac{\alpha}{2}\hbar(k+k')\int_{0}^{L} ds \bra{\downarrow} {\bm \sigma}\cdot {\bm e}_{B}^{0}(s) \ket{\uparrow}\sin(\beta s)\frac{e^{i(k-k')s}}{2\pi\hbar}\\
    &=\frac{\alpha}{2}\hbar(k+k')\int_{0}^{L} ds \frac{-i P_0 e^{is/D}}{D}\sin(\beta s)\frac{e^{i(k-k')s}}{2\pi\hbar},
\end{align}
where $p_s \ket{k}=\hbar k \ket{k}$ and $\bra{\downarrow} {\bm \sigma}\cdot {\bm e}_{B}^{0}(s) \ket{\uparrow}=-i (P_0/D) e^{is/D}$. When the length $L$ of a helix is sufficiently large, the transition amplitude becomes non-negligible only if $|D^{-1}\pm\beta+k-k'|$ is close to zero (rotating wave approximation), namely $|\Delta k + (D^{-1}\pm\beta)|\approx 0$ with $\Delta k = k - k'$, as discussed in the main manuscript. Here $\pm \beta$ term originates from $\sin(\beta s)= (e^{i\beta s}-e^{-i\beta s})/(2i)$. Note that the spin-dependent momentum conservation condition is invariant even if ${\bm e}_{T}^{0}(s)$ is considered instead of ${\bm e}_{B}^{0}(s)$, as $\bra{\downarrow}{\bm e}_{T}^{0}(s)\ket{\uparrow}=i(R_{0}/D)e^{is/D} \propto e^{is/D}$. The same holds for the linear combinations of ${\bm e}_{B/T}^{0}(s)$, describing breathing and longitudinal modes.

\section{Numerics}

When an electron wavepacket is coupled to a single phonon mode, the total state is described by a $2M$-component spinor wavefunction with $M$ denoting the number of phonon number states considered in simulations
\begin{align} \label{SI:eq:spinor}
    [a,b]\ni s\mapsto (\psi^0_\uparrow(s),\psi^0_\downarrow(s),\psi^1_\uparrow(s),\psi^1_\downarrow(s),\cdots,\psi^{M-1}_\uparrow(s),\psi^{M-1}_\downarrow(s)).
\end{align} 
The interval $[a,b]$ is divided into three regions $[a,0)$, $[0,L]$ and $(L,b]$ corresponding to input electrode, single-stranded helix and output electrode, respectively. The initial Gaussian electron wavepacket is modeled by $\psi(s)\propto \exp(-(s-\eta)^{2}/(2\zeta^2)+ik_0 s)$ where the width is taken to be $\zeta=25/k_0$ for $(\hbar k_0)^2/(2m)=1\,{\rm eV}$ and the center $\eta$ is chosen in such a way that the initial state enters the helix from the input electrode. We assume that the electron and phonon mode are initially uncorrelated. As discussed in the main manuscript, the phonon state is taken to be vacuum state, coherent state or thermal state, depending on the physical situations considered in simulations. The spin state is assumed to be a random mixture of spin up and down states.

As discussed in the main manuscript, the Hamiltonian is modeled by
\begin{align} \label{SI:eq:ham}
    H(s) = \frac{p_s^2}{2m} + \Pi (s)H_{d},\qquad \Pi(s) =
    \begin{cases}
    1 & s\in [0,L], \\
    0 & \text{else},
    \end{cases}
\end{align}
where $H_d$ denotes a dynamic spin-orbit coupling. In order to avoid an unrealistic sudden change in Hamiltonian, we consider a smoothstep function
\begin{align}\label{SI:eq:smooth}
    f_{s_0,l}(s) = 6 x_{s_0,l}^{5}(s) - 15 x_{s_0,l}^{4}(s) + 10 x_{s_0,l}^{3}(s),\quad x_{s_0,l}(s)=\frac{s-(s_0-l/2)}{l},
\end{align}
which gradually changes from 0 to 1 over $s_0-l/2\le s\le s_0+l/2$. The smoothstep function is applied to the rectangular function $\Pi(s)$ at $s_0=0$ and $L$ with $l=1\,{\rm nm}$, so that the Hamiltonian changes gradually at the interfaces between electrodes and helix.

To compute wavepacket dynamics numerically, we employ the fourth-order Runge-Kutta method. The interval $[a,b]$ is divided into $N$ discrete points $(s_1,\cdots,s_N)$ with uniform $\Delta s = s_{j+1}-s_j$. Accordingly, the hybrid electron-phonon state $\ket{\psi}$ is described by a $2MN$-dimensional vector. In simulations, $N$ is increased until wavepacket dynamics shows numerical convergence ($\Delta s\sim \SI{0.01}{\nano\meter}$). The time interval $[0,t]$ is divided into $K$ time points with $\Delta t=t/K$ and $K$ is increased until simulated results show numerical convergence ($\Delta t \sim\SI{0.01}{\femto\second}$).

The electron momentum operator and kinetic energy, proportional to $\partial_s$ and $\partial_s^2$, respectively, are described by the fourth-order finite difference method
\begin{align*}
    \eval{\pdv{\psi}{s}}_{s=s_j} &\approx \frac{1}{\Delta s}\left(\frac{1}{12}\psi_{j-2} -\frac{2}{3}\psi_{j-1} + \frac{2}{3}\psi_{j+1}-\frac{1}{12}\psi_{j+2}\right),\\
    \eval{\pdv[2]{\psi}{s}}_{s=s_j} &\approx \frac{1}{\Delta s^2}\left(-\frac{1}{12}\psi_{j-2} +\frac{4}{3}\psi_{j-1} - \frac{5}{2}\psi_j + \frac{4}{3}\psi_{j+1}-\frac{1}{12}\psi_{j+2}\right).
\end{align*}

\section{Markovian Closure}

In simulations, we consider an initial Gaussian electron wavepacket propagating from input electrode, via a single-stranded helix, to output electrode. The width of the electron wavepacket determines the standard deviation of its momentum distribution. In case that the initial state has a well-defined momentum, where its average value is larger than its standard deviation, the initial Gaussian state becomes significantly broad and as a result one needs to consider a long input electrode, increasing the dimension of the Hamiltonian. In addition, as the input and output electrodes are truncated in simulations, the reflection occurred at the ends of the electrodes can make the electron wavepacket re-enter the helix. Such a finite size effect can introduce numerical errors in the computation of spin polarization unless the lengths of the electrodes and simulation times are chosen appropriately.

To avoid the finite size effect caused by truncated electrodes, we attach a sink to the end of each electrode, so that the reflection of the electron wavepacket at the end point is significantly suppressed. The reflection probability can be reduced to less than $\sim 0.1\,\%$ by optimizing the structure and  parameters of the sink. As discussed in the main manuscript and the references therein, the sink, which we call Markovian closure, can be modeled by a network of ten sites under Lindbladian damping, which are coupled to each end point of an electrode conditional to spin and phonon states. It is found that the values of the spin polarization computed with and without the Markovian closure are quantitatively well matched, while the Markovian closure enables one to reduce the length of the output electrode significantly, making numerical simulations more efficient.

\section{Quantum-Jump Algorithm}

In the main manuscript, we consider a phonon relaxation noise modeled by the Lindblad equation
\begin{align}
    \partial_t \rho(t) &= -(i/\hbar)[H,\rho(t)] +\gamma n(a^{\dagger}\rho(t)a-\{a a^{\dagger}\!,\rho(t)\}/2)+ \gamma (1+n)(a\rho(t)a^{\dagger}-\{a^{\dagger}a,\rho(t)\}/2).
\end{align}
Since a continuous variable electron wavepacket is considered in our work, where its spatial resolution is increased until wavepacket dynamics shows numerical convergence, the simulations become challenging when a density matrix $\rho(t)$ is considered instead of a pure state. Such a Lindblad noise can be taken into account in the simulations of pure state dynamics by using a quantum-jump algorithm, as discussed in the main manuscript and references therein.

In more details, for a normalized pure state $\ket{\psi(t)}$ at time $t$, the probabilities for creating and annihilating a single phonon between times $t$ and $t+dt$ are defined, respectively, by
\begin{align}
    P_{c}&=\gamma n {\rm tr}[a^{\dagger}\ket{\psi(t)}\bra{\psi(t)}a]dt,\\
    P_{a}&=\gamma (1+n) {\rm tr}[a\ket{\psi(t)}\bra{\psi(t)}a^{\dagger}]dt,
\end{align}
with ${\rm tr}$ denoting the trace. In the quantum-jump algorithm, a random number $x$ is generated from a uniform distribution between 0 and 1. When the random number $x$ satisfies $0\le x< P_a$, we assume that a single phonon is annihilated, described by $\ket{\psi(t+dt)}=a\ket{\psi(t)}/\sqrt{\bra{\psi(t)}a^{\dagger}a\ket{\psi(t)}}$. When the random number $x$ satisfies $P_a\le x< P_a+P_c$, we assume that a single phonon is created, described by $\ket{\psi(t+dt)}=a^{\dagger}\ket{\psi(t)}/\sqrt{\bra{\psi(t)}aa^{\dagger}\ket{\psi(t)}}$. Otherwise we assume that no quantum jump occurs and the time evolution of the pure state $\ket{\psi(t)}$ under the phonon relaxation noise is governed by a non-Hermitian Hamiltonian $H'=H-i\hbar\gamma n a a^{\dagger}/2-i\hbar\gamma (1+n) a^{\dagger} a/2$, leading to $\ket{\psi(t+dt)}=\exp(-(i/\hbar)H'dt)\ket{\psi(t)}$. Here the anti-commutator terms from the Lindblad equation make the norm of the time-evolved state $\ket{\psi(t+dt)}$ smaller than unity, so it should be normalized at every time step. It can be shown that an ensemble average of the pure state dynamics under quantum jumps is equivalent to the density matrix dynamics obtained by solving the full Lindblad equation.

\end{widetext}

\end{document}